\documentclass[twocolumn,floatfix,showpacs,prb,eqsecnum]{revtex4}
\usepackage{graphicx}
\usepackage{amsmath}
\usepackage{amssymb}
\usepackage{bm}

\usepackage{color}

\begin{document}

\title{Majorana fermions emerging from magnetic nanoparticles on a superconductor without spin-orbit coupling}
\author{T.-P. Choy}
\affiliation{Instituut-Lorentz, Universiteit Leiden, P.O. Box 9506, 2300 RA Leiden, The Netherlands}
\author{J. M. Edge}
\affiliation{Instituut-Lorentz, Universiteit Leiden, P.O. Box 9506, 2300 RA Leiden, The Netherlands}
\author{A. R. Akhmerov}
\affiliation{Instituut-Lorentz, Universiteit Leiden, P.O. Box 9506, 2300 RA Leiden, The Netherlands}
\author{C. W. J. Beenakker}
\affiliation{Instituut-Lorentz, Universiteit Leiden, P.O. Box 9506, 2300 RA Leiden, The Netherlands}
\date{August 2011}

\begin{abstract}
There exists a variety of proposals to transform a conventional \textit{s}-wave superconductor into a topological superconductor, supporting Majorana fermion mid-gap states. A necessary ingredient of these proposals is strong spin-orbit coupling. Here we propose an alternative system consisting of a one-dimensional chain of magnetic nanoparticles on a superconducting substrate. No spin-orbit coupling in the superconductor is needed. We calculate the topological quantum number of a chain of finite length, including the competing effects of disorder in the orientation of the magnetic moments and in the hopping energies, to identify the transition into the topologically nontrivial state (with Majorana fermions at the end points of the chain).
\end{abstract}
\pacs{74.45.+c, 74.78.Na, 75.30.Hx, 75.75.Lf}

\maketitle

\section{Introduction}
\label{intro}

The search for Majorana fermions in superconducting nanowires is both rewarding and difficult.\cite{Wil09,Fra10,Ser11} These nondegenerate midgap states at the end points of the wire require that the \textit{s}-wave proximity effect coexists with broken time-reversal and spin-rotation symmetries, which together can drive the superconductor into a topologically nontrivial phase.\cite{Kitaev01} While the former symmetry can be readily broken by a magnetic field, it has been argued that Rashba spin-orbit coupling is too weak to effectively break the latter symmetry.\cite{Pot11} We are optimistic that some variation on the InAs nanowire proposal\cite{Lut10,Ore10} will be successfully realized, but alternative proposals\cite{Fu08,Lee09,Pot10,Duc11,chung,Wen11} continue to play an important role.

\begin{figure}[tb]
\centerline{\includegraphics[width=0.8\linewidth]{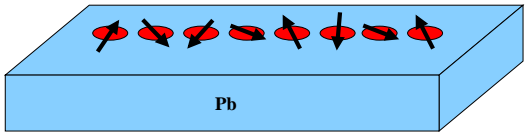}}
\caption{\label{setup}
Proposed setup to obtain Majorana fermions at the end points of a chain of magnetic nanoparticles (typically Fe or Ni) on an \textit{s}-wave superconducting substrate (typically Pb). Arrows indicate the orientation of magnetic moments of the nanoparticles.
}
\end{figure}

In this paper we propose a route to Majorana fermions in \textit{s}-wave superconductors that does not at all require materials with spin-orbit coupling. We consider a one-dimensional chain of magnetic nanoparticles (magnetic dots\cite{Martin}) on a superconducting substrate. (See Fig.\ \ref{setup}.) The nanoparticles create bound states in the superconducting gap, having a nonzero magnetic moment.\cite{Yu,shiba,rusinov} The magnetic moment breaks time-reversal symmetry as well as spin-rotation symmetry, without the need for spin-orbit coupling in the superconductor.

As we will show, the transition into the topologically nontrivial phase is governed by the competition of two types of disorder: On the one hand, disorder in the orientation of the magnetic moments on nearby nanoparticles is needed to open a gap in the excitation spectrum --- which is a prerequisite for a topological phase. On the other hand, disorder in the hopping energies localizes the states on short segments of the chain --- suppressing the superconducting order needed for the topological phase.

In the next section we introduce the model of a chain of magnetic nanoparticles on a superconductor, and then in Sec.\ \ref{sec:scattering} we use scattering theory\cite{Akhmerov1} to calculate the topological quantum number of a finite chain, including the competing effects of magnetic and hopping disorder. As shown in Sec.\ \ref{phasediagram}, we find a substantial region in parameter space that is topologically nontrivial (and thus has Majorana fermions at the end points of the chain).

\section{Model of magnetic nanoparticles on a superconductor}
\label{model}

\subsection{Hamiltonian}
\label{modelSnp}

For each magnetic nanoparticle on top of the superconductor we consider a single electronic orbital near the Fermi level $\mu$, with spin $\alpha$ coupled to the local magnetic moment through an effective magnetic or exchange field $\bm{B}_{n}$. We set $\hbar$ and the Zeeman energy $g\mu_{\rm B}$ both equal to unity, so $\bm{B}_{n}$ is measured in units of energy. We assume that the magnetic moment has the same magnitude $B_{0}=|\bm{B}_{n}|$ on each nanoparticle, varying only in the orientation. States on neighboring nanoparticles $n,n+1$ are coupled by a hopping energy $t_{n}$, where the $n$-dependence accounts for variations in the nanoparticle spacing.

The superconducting substrate induces a spin-singlet pairing energy $\Delta_{0}$ on the nanoparticles, taken as $n$-independent. The electrostatic potential induced by the superconductor is also taken as $n$-independent, so that it simply gives an offset to $\mu$. The charging energy $e^{2}/C$ is unimportant because the nanoparticle is strongly coupled to the superconductor.

The Hamiltonian of this model is given by
\begin{align}
H={}&  \sum_{n,\alpha} \left(t_{n} f^\dagger_{n\alpha} f_{n+1,\alpha}+\text{H.c}\right) - \mu \sum_{n,\alpha} f^\dagger_{n\alpha} f_{n\alpha}\nonumber\\
&+ \sum_{n,\alpha,\beta} (\bm{B}_n \cdot\bm{\sigma})_{\alpha\beta} f^\dagger_{n\alpha}  f_{n\beta}\nonumber\\
&+ \sum_{n} \left(\Delta_{0}f^\dagger_{n\uparrow} f^\dagger_{n\downarrow} + \text{H.c.} \right),
\label{h0}
\end{align}
where the abbreviation H.c.\ stands for Hermitian conjugate. The operator $f_{n\alpha}$ is the fermion operator for a spin-$\alpha$ electron on the $n$-th nanoparticle and $\bm{\sigma}=(\sigma_{x},\sigma_{y},\sigma_{z})$ denotes the vector of Pauli matrices. The magnetic moments are taken as frozen, without any dynamics of their own.

Upon transformation to the Bogoliubov basis $\Psi_n =(f_{n\uparrow},f_{n\downarrow},f_{n\downarrow}^\dagger,-f_{n\uparrow}^\dagger)$, the Hamiltonian takes the form
\begin{equation}
 H = \sum_n \left[\left( \Psi_n^\dagger
\hat t_n \Psi_{n+1} + \text{H.c.} \right) + \Psi_n^\dagger \hat h_n \Psi_n
\right].\label{eq:H-compact}
\end{equation}
The matrices $\hat h_n$ and $\hat t_n$ are defined by
\begin{align}
\hat h_n ={}& \left( \begin{array}{cc}
-\mu\sigma_0 + \bm{B}_n \cdot \bm{\sigma} & \Delta_{0}\sigma_0\\
\Delta_{0}\sigma_0 & \mu\sigma_0 + \bm{B}_n \cdot \bm{\sigma}
\end{array} \right)\nonumber\\
={}& -\mu\sigma_0 \tau_z + (\bm{B}_n \cdot \bm{\sigma}) \tau_0 + \Delta_{0} \sigma_0\tau_x,\label{eq:hn}\\
\hat t_n ={}& \left( \begin{array}{cc}
t_{n}\sigma_0 & 0\\
0 & -t_{n} \sigma_0
\end{array} \right)= t_{n} \sigma_0 \tau_z. \label{eq:tn}
\end{align}
The Pauli matrices $\sigma_i$ and $\tau_i$ ($i=x,y,z$), with identity matrices $\sigma_0$ and $\tau_0$, act on the spin space and particle-hole space, respectively.

The Hamiltonian \eqref{eq:H-compact} has neither time-reversal nor spin-rotation symmetry, but it does satisfy the particle-hole symmetry requirement that
\begin{equation}
H\mapsto-H\;\;{\rm for}\;\;\Psi_{n}\mapsto \sigma_{y}\tau_{y}\Psi_{n}^{\dagger}.\label{ehsymmetry}
\end{equation}
This places the system in symmetry class D, which in one dimension has a topologically nontrivial phase.\cite{Ryu10}

\subsection{Single-band limit}
\label{singelband}

In the large magnetization regime $B_{0}\simeq|\mu|\gg \Delta_{0},t_{n}$ the electron spin on the $n$-th nanoparticle is nearly polarized along $\bm{B}_{n}$. As outlined in App.\ \ref{sec:rotation}, the Hamiltonian can then be projected onto the lowest spin band, with electron operator $\psi_{n}$. Only virtual transitions to the higher spin band contribute. To first order in $1/B_{0}$ the effective Hamiltonian has the form
\begin{align}
H_{\rm{eff}} ={}& \sum_n \biggl[ \bigl( \tilde{t}_{n} \psi^\dagger_n \psi_{n+1}+\tilde{t}'_{n} \psi^\dagger_n \psi_{n+2}  + \text{H.c.} \bigr) - \tilde{\mu}_{n} \psi^\dagger_n \psi_n \nonumber\\
&+ \bigl( \tilde{\Delta}_n  \psi^{\dagger}_{n} \psi^{\dagger}_{n+1} + \text{H.c.} \bigr) \biggr],\label{proj}
\end{align}
with coefficients defined in App.\ \ref{sec:rotation}. The effective pair potential $\tilde{\Delta}_{n}$ is of order $\Delta_{0}t_{n}/B_{0}$, dependent on the relative angle between $\bm{B}_{n}$ and $\bm{B}_{n+1}$. For parallel magnetic moments $\tilde{\Delta}_{n}$ vanishes.

The single-band Hamiltonian \eqref{proj} has the same form as Kitaev's spinless \textit{p}-wave superconducting chain.\cite{Kitaev01} The difference is that here the \textit{p}-wave pairing is obtained from \textit{s}-wave pairing due to the coupling of the electron spin to local magnetic moments. This has the effect of coupling spin to orbital degrees of freedom, but in contrast to existing proposals,\cite{Lut10,Ore10,Fu08,Lee09,Pot10,Duc11,chung,Wen11} neither spin-orbit coupling in the superconductor is needed nor a Rashba effect in the nanowire.

We have given the single-band limit \eqref{proj} to make contact with Kitaev's model. In what follows we will use the full Hamiltonian \eqref{eq:H-compact}, valid to all orders in $1/B_{0}$.

\subsection{Disorder}
\label{disorder}

We distinguish the localizing effect of disorder in the hopping energy $t_{n}$, which localizes the electrons without opening an
excitation gap, from the gap opening effect of disorder in the orientation $\bm{b}_{n}=\bm{B}_{n}/B_{0}$ of the magnetic moments.

Disorder in the hopping energies, due to variations in the interparticle spacing, is modeled by drawing the $t_{n}$'s ($n=1,2,\ldots N-1$) independently from a uniform distribution in the interval $(t_{0}-\delta_{t}t_{0}, t_{0}+\delta_{t}t_{0})$.

For the magnetic moments we take a dipolar ferromagnetic correlation of the unit vectors $\bm{b}_{n}$ on neighboring nanoparticles, according to the distribution
\begin{equation}
P(\bm{b}_{1},\bm{b}_{2},\ldots \bm{b}_{N})=(4\pi)^{-N}\prod_{n=1}^{N-1}\frac{\exp\left(\bm{b}_{n}\cdot\bm{b}_{n+1}/\delta_b\right)}{\delta_{b}\sinh\delta_{b}^{-1}},\label{Pbdef}
\end{equation}
where the parameter $\delta_{b}>0$ quantifies the strength of the correlation (strongly correlated for small $\delta_{b}$).

\section{Scattering matrix}
\label{sec:scattering}

To identify the topologically nontrivial phase of a finite disordered chain of $N$ nanoparticles it is more efficient to work with the scattering matrix than the Hamiltonian.\cite{Akhmerov1} The scattering matrix ${\cal S}$ relates incoming and outgoing wave amplitudes at the Fermi level. The waves can come in from the left end or from the right end of the chain, in two spin directions and as electron or hole, so ${\cal S}$ is an $8\times 8$ unitary matrix. Its $4\times 4$ sub-blocks are the reflection and transmission matrices,
\begin{equation}
{\cal S}=\begin{pmatrix}
{\cal R}&{\cal T}'\\
{\cal T}&{\cal R}'
\end{pmatrix}.\label{Sdef}
\end{equation}
Particle-hole symmetry \eqref{ehsymmetry} requires that
\begin{equation}
{\cal X}=\sigma_{y}\tau_{y}{\cal X}^{\ast}\sigma_{y}\tau_{y},\;\;{\cal X}\in\{{\cal R},{\cal R}',{\cal T},{\cal T}'\}.\label{Xehsym}
\end{equation}

Following Ref.\ \onlinecite{Ful11} we calculate the scattering matrix by writing the tight-binding equations at the Fermi level in the form
\begin{subequations}
\label{PhiMn}
\begin{align}
&\begin{pmatrix}
\hat{t}_{n}\Phi_{n}\\
\Phi_{n+1}
\end{pmatrix}={\cal M}_{n}
\begin{pmatrix}
\hat{t}_{n-1}\Phi_{n-1}\\
\Phi_{n}
\end{pmatrix},\label{PhiMna}\\
&{\cal M}_{n}=\begin{pmatrix}
0&\hat{t}_{n}\\
-\hat{t}_{n}^{-1}&-\hat{t}_{n}^{-1}\hat{h}_{n}
\end{pmatrix},\label{PhiMnb}
\end{align}
\end{subequations}
where $\Phi_{n}$ is a four-component vector of wave amplitudes on site $n$. (Sites $n=0$ and $n=N+1$ represent electron reservoirs.) Waves at the two ends of the chain are related by the transfer matrix\cite{note1}
\begin{equation}
{\cal M}={\cal M}_{N}{\cal M}_{N-1}\cdots{\cal M}_{2}{\cal M}_{1}.\label{calMdef}
\end{equation}
Current conservation is expressed by the identity
\begin{equation}
{\cal M}^{\dagger}_{n}\Sigma_{y}{\cal M}_{n}=\Sigma_{y},\;\;\Sigma_{y}=\begin{pmatrix}
0&-i\\
i&0
\end{pmatrix}.\label{MdaggerSigmayM}
\end{equation}

We transform to a new basis with right-moving and left-moving waves separated in the upper and lower four components, by means of the unitary transformation
\begin{equation}
\tilde{\cal M}={\cal U}^{\dagger}{\cal M}{\cal U},\;\;
{\cal U}=\sqrt{\frac{1}{2}}\begin{pmatrix}
1&1\\
i&-i
\end{pmatrix}.\label{JU}
\end{equation}
The current conservation relation \eqref{MdaggerSigmayM} transforms to
\begin{equation}
\tilde{\cal M}^{\dagger}_{n}\Sigma_{z}\tilde{\cal M}_{n}=\Sigma_{z},\;\;\Sigma_{z}=\begin{pmatrix}
1&0\\
0&-1
\end{pmatrix}.\label{MdaggerSigmazM}
\end{equation}
In this basis the transmission and reflection matrices follow from
\begin{equation}
\begin{pmatrix}
{\cal T}\\
0
\end{pmatrix}=\tilde{\cal M}\begin{pmatrix}
1\\
{\cal R}\end{pmatrix},\;\;
\begin{pmatrix}
{\cal R}'\\
1
\end{pmatrix}=\tilde{\cal M}\begin{pmatrix}
0\\
{\cal T}'
\end{pmatrix}.\label{TRdef}
\end{equation}

Unitarity together with particle-hole symmetry \eqref{Xehsym} ensure that the determinants of ${\cal R}$ and ${\cal R}'$ are identical real numbers. The $\mathbb{Z}_{2}$ topological quantum number ${\cal Q}=\pm 1$ is given by\cite{Akhmerov1,Mer02}
\begin{equation}
{\cal Q}={\rm sign}\,{\rm Det}\,{\cal R}.\label{QRrelation}
\end{equation}
As shown in Ref.\ \onlinecite{Akhmerov1}, Majorana bound states exist at the end points of the chain if and only if ${\cal Q}=-1$, so this identifies the topologically nontrivial phase.

\section{Phase diagram}
\label{phasediagram}

\begin{figure}[tb]
\centerline{\includegraphics[width=0.7\linewidth]{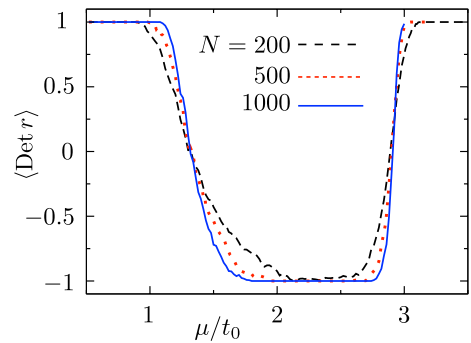}}
\caption{\label{fig_ensemble}
Determinant of the reflection matrix as a function of $\mu$, for fixed $B_{0}=2t_{0}$ and $\Delta=0.9\,t_{0}$, ensemble averaged over $400$ chains of nanoparticles of  length $N$. This is data for $\delta_{b}=\infty$, $\delta_{t}=0$, corresponding to random and uncorrelated orientations of the magnetic moments of the nanoparticles, without randomness in the hopping energies.
}
\end{figure}

\begin{figure}[tb]
\centerline{\includegraphics[width=0.6\linewidth]{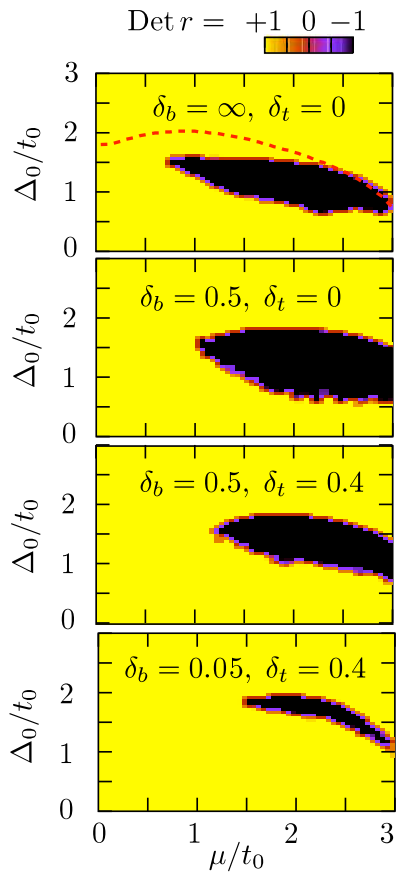}}
\caption{\label{fig_phase}
Color scale plot of the determinant of the reflection matrix as a function of $\mu$ and $\Delta_{0}$, for fixed $B_{0}=2t_{0}$, calculated for a single disorder realization in a chain of $N=6000$ nanoparticles. The topologically nontrivial phase has ${\rm Det}\,r=-1$ (black region). The four panels are for different values of the disorder $\delta_{t}$ in the hopping energies and $\delta_{b}$ in the orientation of the magnetic moments [with $\delta_{b}=\infty$ corresponding to random and uncorrelated orientations, see Eq.\ \eqref{Pbdef}]. The dashed red curve in the top panel gives the phase boundary following from the self-consistent Born approximation.
}
\end{figure}

\begin{figure}[tb]
\centerline{\includegraphics[width=0.6\linewidth]{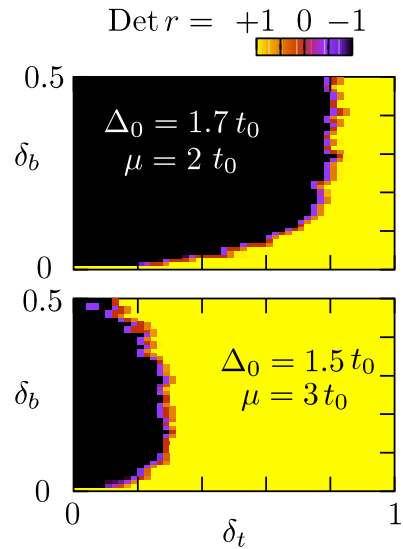}}
\caption{\label{fig_phase_2}
Same as Fig.\ \ref{fig_phase}, but now as a function of $\delta_{b}$ and $\delta_{t}$ for fixed $\Delta_{0},\mu$.
}
\end{figure}

Results of our search for a topologically nontrivial phase are shown in Figs.\ \ref{fig_ensemble}--\ref{fig_phase_2}. We have kept $B_{0}=2t_{0}$ fixed and varied the material parameters $\Delta_{0},\mu$ as well as the disorder parameters $\delta_{b},\delta_{t}$. The color scale Figs.\ \ref{fig_phase} and \ref{fig_phase_2} of ${\rm Det}\,r$ show the topologically nontrivial region emerging as a black island (${\rm Det}\,r=-1$) in a yellow background (${\rm Det}\,r=1$). These are plots for a single disorder realization and a single system size ($N=6000$). The $N$-dependence of the ensemble average $\langle{\rm Det}\,r\rangle$ is plotted in Fig.\ \ref{fig_ensemble}, to show how the transition becomes sharper with increasing system size.

The competing effects of the two types of disorder are evident in the phase diagrams. We see that the topologically nontrivial region is largest for $\delta_{t}$ equal to zero and $\delta_{b}$ small but not too small. Disorder in the hopping energies or in the magnetic moments causes localization, which reduces the nontrivial region (mainly on the small $\Delta_{0}$ side). Some disorder in the orientation of the magnetization is needed to open a gap in the spectrum, so while the optimal value of $\delta_{t}$ equals zero, the optimal value of $\delta_{b}$ is small but nonzero.

To obtain more insight in the $\Delta_{0}$ and $\mu$ dependence of the phase diagram, we have calculated the phase boundary by the method of self-consistent Born approximation ({\sc scba}). As shown in Ref.\ \onlinecite{Groth} for a different system, the {\sc scba} can locate the gap inversion associated with the topologically nontrivial phase. The calculation is described in App.\ \ref{sec:Born} and the result is shown in Fig.\ \ref{fig_phase} (red dashed curve in top panel), for a system without disorder in the hopping energies ($\delta_{t}=0$) and with random and uncorrelated directions of the magnetic moments ($\delta_{b}=\infty$). 

The {\sc scba} describes reasonably well, without any adjustable parameters, the location of the phase boundary at large $\Delta_{0}$. The phase boundary at small $\Delta_{0}$ is not recovered by the {\sc scba}. Our explanation is that the former phase boundary is due to gap inversion, while the latter phase boundary is due to localization (which is beyond the reach of the {\sc scba}).

\section{Conclusion}
\label{conclude}

In conclusion, we have proposed to create Majorana fermions by depositing magnetic nanoparticles on a superconducting substrate. The superconductor has conventional \textit{s}-wave pairing and need not have any spin-orbit coupling. The candidate material for the magnetic nanoparticles is iron or nickle. For the superconductor one could choose lead, with a coherence length $\xi\simeq 80\,{\rm nm}$.

The magnetic moments of nearby nanoparticles (spacing small compared to $\xi$) are correlated by dipolar interactions.\cite{stepanyuk} As we have found, a strong correlation in the orientation of the magnetic moments ($\delta_b\ll 1$) does not preclude the appearance of a topologically nontrivial phase --- only a nearly complete alignment suppresses it. The most stringent requirement seems to be the need to have little disorder in the hopping energies, since in this one-dimensional system the localization length is of the order of the mean free path and hence severely limited by disorder. The spacing of the nanoparticles should therefore be highly uniform.

\begin{figure}[tb]
\centerline{\includegraphics[width=0.7\linewidth]{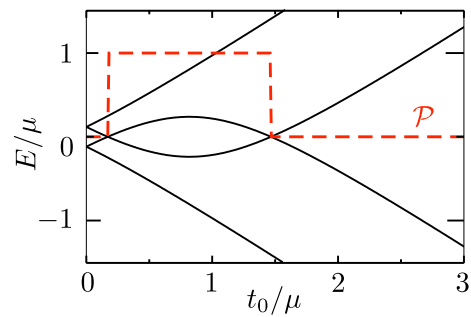}}
\caption{\label{fig_2sites}
Excitation spectrum (black solid curves) of two magnetic particles with an angle $\theta=70^{\circ}$ between their magnetic moments, calculated from the Hamiltonian \eqref{h0} at fixed $\mu=B_{0}=2\Delta_{0}$ as a function of the hopping energy $t_{0}$. The level crossings at the Fermi energy do not split because the ground states at the two sides of the crossing differ in fermion parity ${\cal P}\in\{0,1\}$ [red dashed curve, calculated from the Pfaffian of the antisymmetrized Hamiltonian, ${\cal P}=\frac{1}{2}-\frac{1}{2}\,{\rm sign}\,{\rm Pf}\,(\sigma_{y}\tau_{y}H)$].
}
\end{figure}

While a long chain of nanoparticles (much longer than $\xi$) is needed to create a pair of weakly coupled Majorana fermions at the end points, a signature of a topological phase transition can be identified even if one has only a few nanoparticles. This signature is a protected level crossing at the Fermi level, which signals a change in the fermion parity of the ground state.\cite{Kitaev01} We show this feature in Fig.\ \ref{fig_2sites}, for the simplest case of just two nanoparticles with a non-collinear magnetization. Nanoparticle-induced bound states within the superconducting gap have been resolved by scanning tunneling microscopy, for a Mn dimer adsorbed on a Pb superconducting thin film.\cite{ji} We predict level crossings at the Fermi energy ($E=0$) upon variation of the spacing of the two magnetic atoms, as a signature of the switch from an even to an odd number of electrons in the ground state.

\acknowledgments

We have benefited from discussions with H. Manoharan. This research was supported by the Dutch Science Foundation NWO/FOM, by an ERC Advanced Investigator Grant, and by the EU network NanoCTM.

\appendix
\section{Relationship to Kitaev's model}
\label{sec:rotation}

We establish the relationship between the tight-binding model \eqref{h0} of electrons coupled to an \textit{s}-wave superconducting order parameter and to a randomly oriented magnetization, on the one hand, and Kitaev's model of  a spinless \textit{p}-wave chain,\cite{Kitaev01} on the other hand.

The magnetic moment $\bm{B}_{n}=B_{0}\bm{\hat{b}}_{n}$ on the $n$-th nanoparticle has fixed magnitude $B_{0}$ and orientation $\bm{\hat{b}}_{n}=(\sin\theta_n \cos\phi_n, \sin\theta_n \sin\phi_n, \cos\theta_n)$. We align the spin basis on each site with $\bm{\hat{b}}_{n}$,
\begin{align}
&\begin{pmatrix}
f_{n\uparrow}\\
f_{n\downarrow}
\end{pmatrix}
= U_{n}\begin{pmatrix}
\tilde f_{n\uparrow}\\
\tilde f_{n\downarrow}
\end{pmatrix},\label{tildefdef}\\
&U_{n}=\begin{pmatrix}
\cos (\theta_n/2) & -\sin(\theta_n/2) e^{-i\phi_n} \\
\sin (\theta_n/2) e^{i\phi_n} & \cos(\theta_n/2)
\end{pmatrix}.\label{Undef}
\end{align}
The Hamiltonian \eqref{h0} transforms to
\begin{align}
H={}&  \sum_{n,\alpha\beta} \left(t_{n} \Omega_{n,\alpha\beta} \tilde{f}^\dagger_{n\alpha}\tilde{f}_{n+1,\beta}+\text{H.c}\right) \nonumber\\
&+ \sum_{n,\alpha,\beta} B_{0} \sigma_{z,\alpha\beta} \tilde{f}^\dagger_{n\alpha}  \tilde{f}_{n\beta}- \mu \sum_{n,\alpha} \tilde{f}^\dagger_{n\alpha} \tilde{f}_{n\alpha}\nonumber\\
&+ \sum_{n} \left(\Delta_{0}\tilde{f}^\dagger_{n\uparrow} \tilde{f}^\dagger_{n\downarrow} + \text{H.c.} \right).
\label{h0tildef}
\end{align}
The unitary matrix $\Omega_{n}=U_{n}^{\dagger}U_{n+1}$ has elements
\begin{subequations}
\label{Omegandef}
\begin{align}
&\Omega_{n}=
\begin{pmatrix}
\alpha_n & -\beta_n^* \\
\beta_n & \alpha_n^*
\end{pmatrix},\label{Omegandefa}\\
&\alpha_n =  \cos\frac{\theta_n}{2} \cos\frac{\theta_{n+1}}{2} + \sin\frac{\theta_n}{2} \sin\frac{\theta_{n+1}}{2} e^{-i (\phi_n - \phi_{n+1})},\label{Omegandefb}\\
&\beta_n = -\sin\frac{\theta_n}{2} \cos\frac{\theta_{n+1}}{2} e^{i\phi_n} + \cos\frac{\theta_n}{2} \sin\frac{\theta_{n+1}}{2} e^{i\phi_{n+1}}.\label{Omegandefc}
\end{align}
\end{subequations}

In the regime $B_0\simeq |\mu| \gg t_n, \Delta$, only one of the two spin bands lies near the Fermi level. (For definiteness, we take $\mu<0$, so that the spin-down band is the band near the Fermi level.) The effective low-energy Hamiltonian $H_{\rm eff}$ has only virtual spin-flip transitions, without any matrix elements between the spin bands. We obtain $H_{\rm eff}$ from $H$ by a canonical transformation $H\mapsto e^{-i S} H e^{i S}$ followed by a projection onto the spin-down band.\cite{Fol50,schrieffer,Che04} The Hermitian operator $S$ is expanded as $S = S_1 + S_2 + \cdots$, with $S_n$ of order $t_0^p \Delta_0^q B_0^{-r} \mu^{-s}$ and $p+q=r+s=n$. The term $S_{n}$ is chosen such that it eliminates the $n$-th order matrix elements between the spin bands.

To first order one has
\begin{align}
H\mapsto{}&H+[H,iS_{1}],\label{HS1}\\
S_1 ={}& i\sum_n \biggl[  \frac{t_n\beta_n^*}{2B_0} \left(\tilde{f}_{n+1\uparrow}^\dagger \tilde{f}_{n\downarrow} - \tilde{f}_{n\uparrow}^\dagger \tilde{f}_{n+1\downarrow}\right)\nonumber\\
& +  \frac{\Delta_0}{2\mu} \tilde{f}_{n\downarrow} \tilde{f}_{n\uparrow} - \text{H.c.}  \biggr].\label{S}
\end{align}
The resulting effective single-band Hamiltonian is
\begin{align}
H_{\rm eff} ={}& -\sum_n \left(B_0+\mu+ \frac{|\beta_n t_n|^2 + |\beta_{n-1} t_{n-1}|^2}{2B_0}\right) \psi_n^\dagger \psi_n\nonumber\\
& +\sum_{n}  \biggl[ t_n\alpha_n^* \psi_n^\dagger \psi_{n+1}  + \frac{ t_{n+1} \beta_{n+1}^* t_n\beta_n}{2B_0} \psi_n^\dagger \psi_{n+2} \nonumber\\
& +   \left(\frac{1}{2B_0} - \frac{1}{2\mu}  \right) \Delta_0 t_n \beta_n  \psi_n \psi_{n+1} + \text{H.c.} \biggr],\label{heff}
\end{align}
where we have abbreviated $\psi_{n}=\tilde{f}_{n\downarrow}$. This is the \textit{p}-wave chain model of Kitaev,\cite{Kitaev01} with the addition of a next-nearest-neighbor hopping term. The pair potential $\propto \Delta_{0} t_n\beta_n$ vanishes if the magnetic moments are aligned, because then $\beta_{n}=0$.

\section{Phase boundary in self-consistent Born approximation}
\label{sec:Born}

We calculate the phase boundary of the topologically nontrivial phase using the self-consistent Born approximation ({\sc scba}), starting from the Hamiltonian \eqref{h0tildef} in the locally rotated spin basis. For simplicity we take site-independent hopping energies $t_{0}$ and random, uncorrelated magnetic moments.

On each site $n$ we define four real random variables
\begin{align}
&\varphi_1(n),\varphi_2(n),\varphi_3(n),\varphi_4(n)\nonumber\\
&\quad ={\rm Re}\,\alpha_n - \tfrac{4}{9}, \;{\rm Im}\,\alpha_n,{\rm Re}\,\beta_n,{\rm Im}\,\beta_n,\label{varphidef}
\end{align}
of zero average and variance
\begin{align}
& \langle
\varphi_j (n) \varphi_{j'} (n') \rangle= \delta_{jj'} \delta_{nn'}
\frac{1}{m_j},\label{varphicorr}\\
&(m_1,m_2,m_3,m_4 )= (\tfrac{8}{3},8,4,4).\label{mjdef}
\end{align}
We neglect higher order cumulants, approximating the distribution of the $\varphi_{j}(n)$'s by a Gaussian. (These random variables can be thought of as bosonic fields of mass $m_{j}$.)

We work in the Bogoliubov basis $\Psi_n = (\tilde{f}_{n\uparrow},\tilde{f}_{n\downarrow},\tilde{f}_{n\downarrow}^\dagger,-\tilde{f}_{n\uparrow}^\dagger)$ and transform variables from site indices $n\in\{1,2,\ldots N\}$ to dimensionless momenta $k\in(-\pi,\pi)$. Substitution of Eq.\ \eqref{h0tildef} gives
\begin{align}
&H =\sum_{k} \Psi_{k}^{\dagger} H_{0}(k)\Psi_k+\sum_{j=1}^{4}\sum_{k,q} \varphi_j(k-q)
\Psi_k^\dagger v_{j}(k,q)\Psi_q,\label{HH0Vdef}\\
&H_0(k) =  (\tfrac{8}{9} t_{0}\cos k-\mu)\sigma_{0}\tau_{z}
+ B_0\sigma_{z}\tau_{0} +\Delta_{0} \sigma_{0}\tau_{x}.\label{eq:H_boson}
\end{align}
We have abbreviated
\begin{equation}
\sum_k\equiv \frac{N}{2\pi}\int_{-\pi}^{\pi}dk.\label{sumkdef}
\end{equation}
The interaction vertices are given by
\begin{align}
&v_j(k,q) = \tfrac{1}{2}t_{0}\bigl(a_j e^{i q} + a_j^\dagger e^{-i k}\bigr),\label{vjkpdef}\\
&a_{1}=\sigma_0\tau_z,\;\;a_2=i\sigma_z\tau_z,\;\;a_3=-i\sigma_y\tau_z,\;\;a_4=\sigma_x\tau_0.\label{vdef}
\end{align}
They satisfy $v_j^\dagger (k,q) = v_j(q,k)$ and the particle-hole symmetry condition
\begin{equation}
\sigma_y\tau_y v_j(k,q) \sigma_y\tau_y = - v_j^*(-k,-q).\label{vpsym}
\end{equation}

The self-energy $\Sigma(k)$ at the Fermi level is given in {\sc scba} by the integral equation
\begin{equation}
\label{eq:SCBA}
 \Sigma (k) = -\sum_{j=1}^4 \int_{-\pi}^{\pi}\frac{dq}{2\pi m_{j}} v_j(k,q)  [H_0(q)+\Sigma(q)]^{-1} v_j(q,k).
\end{equation}
The effective Hamiltonian
\begin{equation}
H_{\text{\sc scba}}(k)=H_{0}(k)+\Sigma(k)\label{HeffSigma}
\end{equation}
satisfies the particle-hole symmetry relation
\begin{equation}
\sigma_y  \tau_y H_{\text{\sc scba}}(k) \sigma_y  \tau_y = -H_{\text{\sc scba}}^* (-k).\label{Sigmaehsym}
\end{equation}
In contrast to the {\sc scba} calculations in Refs.\ \onlinecite{Groth,Yamakage}, where the interaction vertices are momentum independent, the randomness of the magnetic moments induces momentum-dependent vertices in Eq. (\ref{HH0Vdef}). As a consequence the self-energy $\Sigma(k)$ becomes momentum dependent.

We calculate the topological quantum number ${\cal Q}$ from $H_{\text{\sc scba}}$ by means of Kitaev's Pfaffian formula,\cite{Kitaev01}
\begin{equation}
{\cal Q}  =\text{sign}\,\bigl\{ {\rm Pf}\,\bigl[\sigma_y\tau_y H_{\text{\sc scba}}(0)\bigr]\,{\rm Pf}\,\bigl[\sigma_y\tau_y H_{\text{\sc scba}}(\pi)\bigr]\bigr\}.\label{Pdef}
\end{equation}
The particle-hole symmetry relation \eqref{Sigmaehsym} ensures that the Pfaffian is calculated of an antisymmetric matrix. In the large-$N$ limit this Hamiltonian expression for the topological quantum number is equivalent to the scattering matrix expression \eqref{QRrelation}.\cite{Akhmerov1}

The $4\times 4$ self-energy matrix can be decomposed in terms of 16 combinations of Pauli matrices $\sigma_{\alpha}\tau_{\beta}$,
\begin{equation}
\Sigma(k) = \Sigma_1(k) \sigma_0\tau_z + \Sigma_2(k) \sigma_z\tau_0 + \Sigma_3(k) \sigma_0\tau_x + \cdots.\label{Sigmadecomposed}
\end{equation}
At $k=0,\pi$ only the terms shown are nonzero. These three terms appear in $H_{\text{\sc scba}}(k)$ as a renormalization of the energy scales
\begin{align}
&\tilde{\mu}(k) = \mu - \Sigma_1(k),\;\;
 \tilde{B}(k) = B_0 + \Sigma_2(k),\nonumber\\
&\tilde{\Delta}(k) = \Delta_{0}+ \Sigma_3(k).\label{renormenergy}
\end{align}

Substitution into Eq.\ \eqref{Pdef} gives the topological quantum number
\begin{align}
{\cal Q} ={}&\text{sign}\,\left\{ \left[\tilde B(0)^{2} -  \tilde\varepsilon(0)^2 - \tilde\Delta(0)^2 \right]\right.\nonumber\\
&\times\left. \left[\tilde B(\pi)^{2} - \tilde\varepsilon(\pi)^2 -
\tilde\Delta(\pi)^2 \right] \right\},\label{eq:Kitaev}\\
\tilde\varepsilon(k)={}&\tfrac{8}{9} t_{0}\cos k-\tilde\mu(k).\label{tildeepsdef}
\end{align}
The phase boundary where ${\cal Q}$ changes sign is indicated in Fig.\ \ref{fig_phase} (top panel). This calculation contains no adjustable parameters.

\end{document}